\documentclass[twocolumn,tighten]{aastex63}

\pdfoutput=1 %for arXiv submission
\usepackage{amsmath,amstext,longtable}
\usepackage[T1]{fontenc}
\usepackage{apjfonts} 
\usepackage[figure,figure*]{hypcap}
\received{2021 May 10}
\revised{2021 May 18}
\accepted{2021 May 21}
\submitjournal{\it The Astrophysical Journal Letters}
\shortauthors{Kirkpatrick et al.}
\shorttitle{The Accident}

\usepackage{color,amsmath,longtable}

\begin{document}

\title{The Enigmatic Brown Dwarf WISEA J153429.75$-$104303.3 (aka "The Accident")}

\correspondingauthor{J.\ Davy Kirkpatrick}
\email{davy@ipac.caltech.edu}

\author[0000-0003-4269-260X]{J.\ Davy Kirkpatrick}
\affiliation{IPAC, Mail Code 100-22, Caltech, 1200 E. California Blvd., Pasadena, CA 91125, USA; davy@ipac.caltech.edu}

\author[0000-0001-7519-1700]{Federico Marocco}
\affiliation{IPAC, Mail Code 100-22, Caltech, 1200 E. California Blvd., Pasadena, CA 91125, USA}

\author[0000-0001-7896-5791]{Dan Caselden}
\affiliation{Gigamon Applied Threat Research, 619 Western Avenue, Suite 200, Seattle, WA 98104, USA}

\author[0000-0002-1125-7384]{Aaron M. Meisner}
\affiliation{NSF's National Optical-Infrared Astronomy Research Laboratory, 950 N. Cherry Ave., Tucson, AZ 85719, USA}

\author[0000-0001-6251-0573]{Jacqueline K.\ Faherty}
\affiliation{Department of Astrophysics, American Museum of Natural History, Central Park West at 79th Street, New York, NY 10034, USA}

\author[0000-0002-6294-5937]{Adam C.\ Schneider}
\affiliation{US Naval Observatory, Flagstaff Station, P.O.Box 1149, Flagstaff, AZ 86002, USA}
\affiliation{Department of Physics and Astronomy, George Mason University, MS3F3, 4400 University Drive, Fairfax, VA 22030, USA}

\author[0000-0002-2387-5489]{Marc J. Kuchner}
\affiliation{NASA Goddard Space Flight Center, Exoplanets and Stellar Astrophysics Laboratory, Code 667, Greenbelt, MD 20771, USA}

\author[0000-0003-2478-0120]{S.\ L.\ Casewell}
\affiliation{School of Physics and Astronomy, University of Leicester, University Road, Leicester LE1 7RH, UK}

\author{Christopher R.\ Gelino}
\affiliation{IPAC, Mail Code 100-22, Caltech, 1200 E. California Blvd., Pasadena, CA 91125, USA}

\author[0000-0001-7780-3352]{Michael C.\ Cushing}
\affiliation{The University of Toledo, 2801 West Bancroft Street, Mailstop 111, Toledo, OH 43606, USA}

\author{Peter R.\ Eisenhardt}
\affiliation{Jet Propulsion Laboratory, California Institute of Technology, MS 169-237, 4800 Oak Grove Drive, Pasadena, CA 91109, USA}

\author[0000-0001-5058-1593]{Edward L.\ Wright}
\affiliation{Department of Physics and Astronomy, University of California Los Angeles, 430 Portola Plaza, Box 951547, Los Angeles, CA, 90095-1547, USA}

\author[0000-0003-1785-5550]{Steven D.\ Schurr}
\affiliation{IPAC, Mail Code 100-22, Caltech, 1200 E. California Blvd., Pasadena, CA 91125, USA}

\begin{abstract}

Continued follow-up of WISEA J153429.75$-$104303.3, announced in \cite{meisner2020}, has proven it to have an unusual set of properties. New imaging data from Keck/MOSFIRE and {\it HST}/WFC3 show that this object is one of the few faint proper motion sources known with  $J-$ch2 $>$ 8 mag, indicating a very cold temperature consistent with the latest known Y dwarfs. Despite this, it has W1$-$W2 and ch1$-$ch2 colors $\sim$1.6 mag bluer than a typical Y dwarf. A new trigonometric parallax measurement from a combination of {\it WISE}, {\it Spitzer}, and {\it HST} astrometry confirms a nearby distance of $16.3^{+1.4}_{-1.2}$ pc and a large transverse velocity of $207.4{\pm}15.9$ km s$^{-1}$. The absolute $J$, W2, and ch2 magnitudes are in line with the coldest known Y dwarfs, despite the highly discrepant W1$-$W2 and ch1$-$ch2 colors. We explore possible reasons for the unique traits of this object and conclude that it is most likely an old, metal-poor brown dwarf and possibly the first Y subdwarf. Given that the object has an {\it HST} F110W magnitude of 24.7 mag, broad-band spectroscopy and photometry from {\it JWST} are the best options for testing this hypothesis.

\end{abstract}


\section{Introduction}

The coldest brown dwarfs are a difficult population to characterize. What little flux they emit is concentrated near 5 $\mu$m, and, because this region is largely unobservable from the ground, space-based missions offer the best chance of discovery and follow-up. Studies at other wavelengths are hampered by intrinsically faint magnitudes. For known Y dwarfs -- the coldest brown dwarfs with effective temperatures below $\sim$450K (\citealt{cushing2011}) -- absolute $J$-band magnitudes range from 19.4 to 28.2 mag (\citealt{kirkpatrick2021}).

Despite these difficulties, the number of known, very cold brown dwarfs is slowly growing, with fifty confirmed or suspected Y dwarfs now cataloged (\citealt{kirkpatrick2021}). Nonetheless, uncovering clear trends in colors and absolute magnitudes is proving elusive. WISE J085510.83$-$071442.5 (hereafter, WISE 0855$-$0714) is the fourth closest (sub)stellar system to the Sun and the coldest brown dwarf known ($\sim$250K; \citealt{luhman2014}). Despite estimates that $15^{+20}_{-11}$ such frigid objects (\citealt{wright2014}) have been imaged in data sets by the {\it Wide-Field Infrared Survey Explorer} ({\it WISE}; \citealt{wright2010, mainzer2014}), it is the only brown dwarf recognized to have $T_{\rm eff} < 350K$ (\citealt{kirkpatrick2021}), although two others may also fall in this temperature zone -- WISEA J083011.95+283716.0 (hereafter, WISE 0830+2837; \citealt{bardalez2020}) and 
WISEPA J182831.08+265037.8 (hereafter, WISE 1828+2650; \citealt{cushing2011}). WISE 1828+2650, in fact, stands out on color-mag plots as 1.5-2.0 magnitudes more luminous at $J$ and ch2 than Y dwarfs of similar $J-$ch2 color (\citealt{kirkpatrick2021}), and it has a unique near-infrared spectrum (\citealt{cushing2011, cushing2021}).

WISE 1828+2650 may be hinting at the variety present among the coldest brown dwarfs, and this diversity is becoming more evident as we obtain detailed characterization of the larger Y dwarf sample. The lack of other discoveries as cold as WISE 0855$-$0714 may imply that it is not a typical member of the ultra-cold brown dwarf population. As discussed in section 9.4 of \cite{kirkpatrick2021}, objects at these coldest temperatures are expected to span a wider range of ages than warmer brown dwarfs, suggesting a wider range of metallicities and photometric properties.

In this {\it Letter}, we present additional follow-up of WISEA J153429.75$-$104303.3 (hereafter, WISE 1534$-$1043), nicknamed "The Accident" by its discoverer D.\ Caselden, who serendipitously spotted it as a high-motion object in a field containing an unrelated motion candidate. New data provide a robust parallax and extremely red $J-$ch2 color. The parallax places it securely within 20 pc, and the colors and absolute magnitudes are unlike those of any brown dwarf known. Thus, WISE 1534$-$1043 appears to be the most enigmatic cold brown dwarf yet identified and bolsters the idea that such objects span a wide range of observational properties. 

\begin{figure}
\includegraphics[scale=0.55,angle=0]{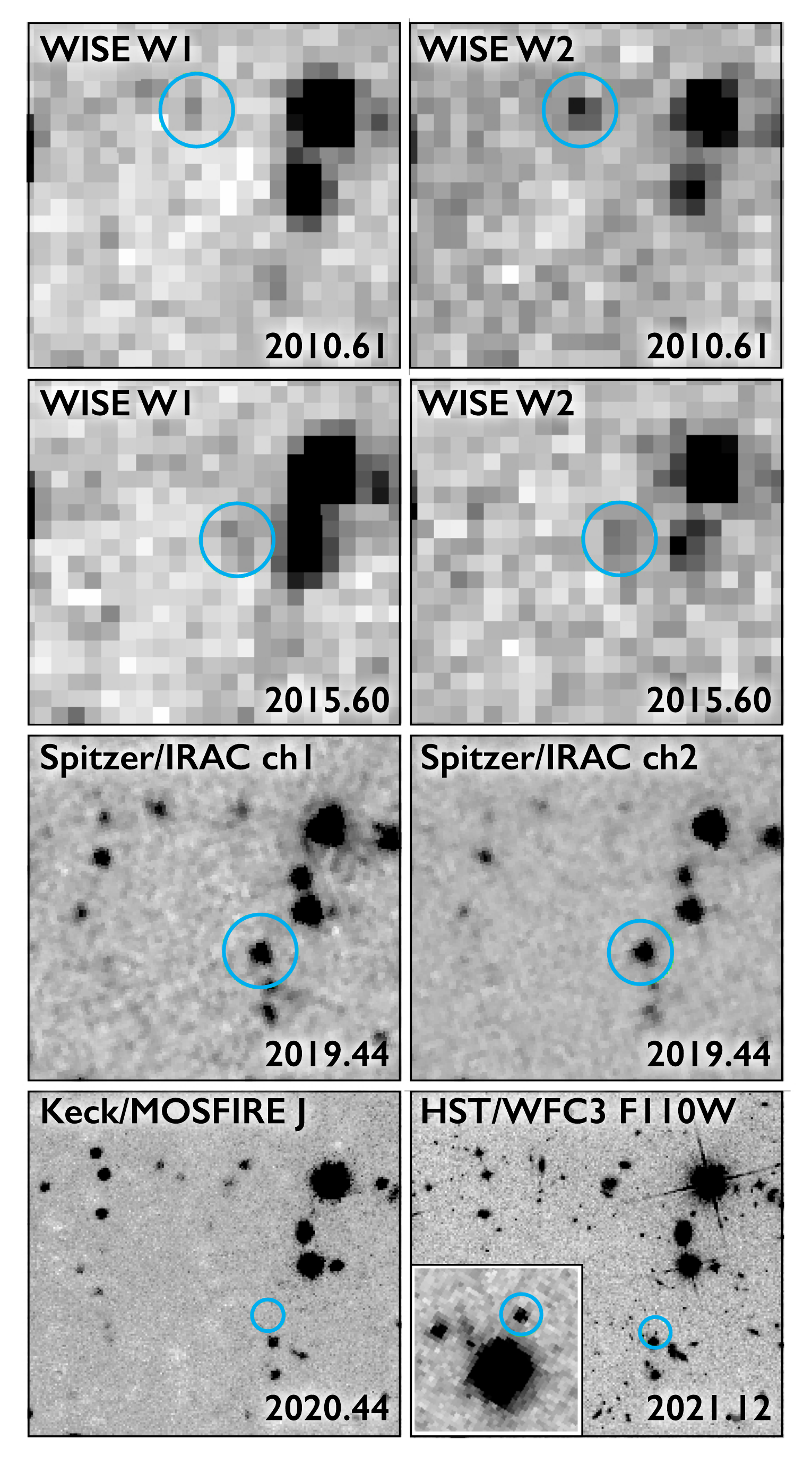}
\caption{Images, each $\sim$1 arcminute on a side, showing the proper motion of WISE 1534$-$1043 (circled in blue). The top two rows show {\it WISE} W1 and W2 images covering a five-year time baseline. The bottom two rows show more recent images from {\it Spitzer}, Keck/MOSFIRE, and HST/WFC3.
The source was not detected in the Keck image, so the position marked is inferred from the astrometric solution. The HST image includes a $5{\arcsec}{\times}5{\arcsec}$ inset highlighting the detection. North is up and east is to the left.
\label{fig:image_sequence}}
\end{figure}

\section{New Data\label{sec:data}}

%Both \cite{meisner2020} and \cite{kirkpatrick2021} have briefly discussed WISE 1534$-$1043 using available {\it WISE} and {\it Spitzer} data. 
\cite{meisner2020} showed that the $J-$ ch2 color limit and presumed high velocity of WISE 1534$-$1043 placed it amongst a small set of mid- to late-T subdwarfs. \cite{kirkpatrick2021} used available {\it WISE} and {\it Spitzer} data to measure a parallax, albeit with a large (30\%) uncertainty. Here we present two new data sets that produce a robust color and parallax reinforcing the uniqueness of this object. 

First, we have obtained a deep $J_{\rm MKO}$-band observation from MOSFIRE (\citealt{mclean2010, mclean2012}) at the W.\ M.\ Keck Observatory. These data, obtained in $\sim1{\farcs}1$ seeing on 2020 Jun 08 (UT) using 60 $\times$ 60s frames in a repeated 5-point dither pattern, were combined with IDL procedures that use bright stars for re-registration and alignment. The result is shown in  Figure~\ref{fig:image_sequence} and compared to earlier unWISE (\citealt{lang2014, meisner2018a, meisner2018b}) and {\it Spitzer} images. Using the preliminary proper motion value from \cite{kirkpatrick2021} to propagate the predicted position to that epoch, we find no detection down to a 3$\sigma$ limit of $J_{\rm MKO}$ = 23.8 mag by calibrating to bright, unsaturated stars whose 2MASS $J$ magnitudes were converted to the MKO system\footnote{ \url{https://irsa.ipac.caltech.edu/data/2MASS/docs/releases/allsky/doc/sec6_4b.html}}. This limit and other pertinent data on WISE 1534$-$1043 are listed in Table~\ref{tab:data}.

Second, we have obtained still deeper imaging at a similar wavelength (F110W) using the Wide Field Camera 3 (WFC3) on the {\it Hubble Space Telescope} ({\it HST}) as part of program 16243 (PI: Marocco). WISE 1534$-$1043 was observed on 2021 Feb 11 (UT) in MULTIACCUM mode with STEP100 sampling for a total exposure time of 2396.92 s. The total exposure was broken into four exposures of 599.23 s each and dithered using the WFC3-IR-DITHER-BOX-MIN pattern, with point spacing of 0$\farcs$572 and line spacing of 0$\farcs$365.

WISE 1534$-$1043 is detected at S/N $\approx$ 5 per exposure with an F110W magnitude of 24.695$\pm$0.083 mag (Table~\ref{tab:data}). As shown in Figure~\ref{fig:image_sequence}, it falls in an uncontaminated region just north of a much brighter object. For astrometry we used the {\it HST} pipeline's calibrated exposures (\textit{flt} files). In each exposure, we measured the ({\it x,y}) positions for all stars via PSF-fitting using the \textit{img2xym} program\footnote{\url{https://www.stsci.edu/~jayander/WFC3/WFC3IR_PSFs/}}, applied the geometric distortion correction provided by STScI, and used the native WCS to compute initial guesses for RA and Dec. To place the observations onto an absolute reference frame, we used these initial positions to find {\it Gaia} Early Data Release 3 (eDR3) sources -- nineteen total -- within the ${\sim}2{\times}2$ arcmin field of view. We used the eDR3 proper motions and parallaxes to update the reference star coordinates from the 2016.0 {\it Gaia} epoch to the 2021.1 {\it HST} epoch. Finally, we determined the standard ({\it x,y}) to (RA,Dec) transformation (see, e.g., equations 1-4 in \citealt{bedin2018}) via least-square fitting. The typical RMS of the fit is 2.5-3.5 mas. The RA and Dec of the nineteen reference stars across the four exposures have a standard deviation of $\sim$1.7 mas.

The ({\it x,y}) position of WISE 1534$-$1043 in each exposure was measured as described above. Due to the relatively low S/N of the detection, its positional measurements have a standard deviation of 22.0 mas in RA and 16.4 mas in Dec. We take these as the empirical uncertainties of our PSF fit, add the RMS of the astrometric calibration in quadrature to compute the RA and Dec uncertainties in each exposure, and take the weighted mean of the four measurements and its uncertainty as the final positional measurement of WISE 1534$-$1043.

\begin{figure*}
\includegraphics[scale=0.875,angle=0]{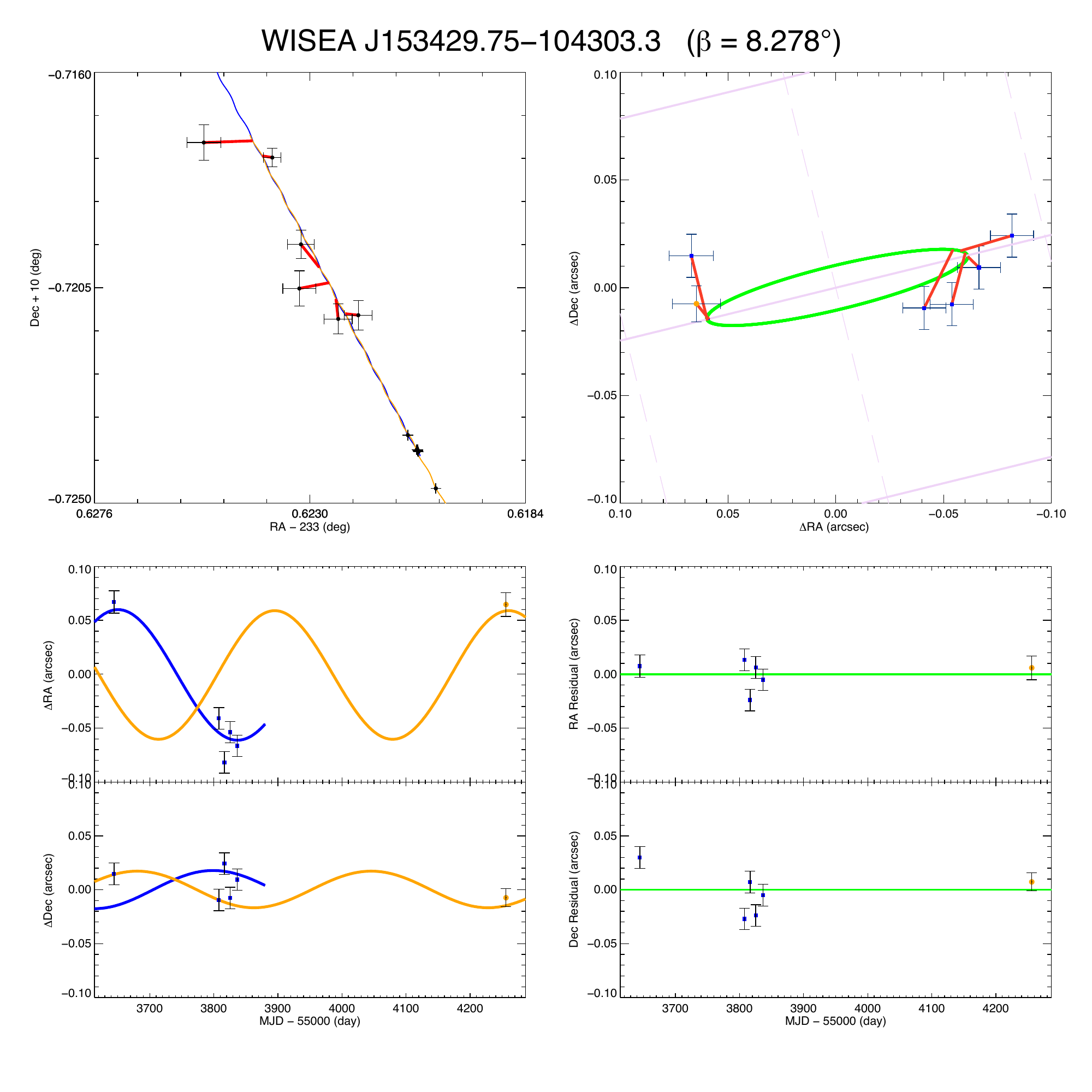}
\caption{Astrometric fit for WISE 1534$-$1043. (Upper left) A square patch of sky showing the measured positions and uncertainties at each epoch (black points with error bars). The blue curve shows the best fit from the vantage point of {\it Spitzer}, and the orange curve shows the best fit from the vantage point of the Earth. Red lines connect each observation to its time on the corresponding best-fit curve. (Upper right) A square patch of sky centered at the mean equatorial position of the target. The green curve is the parallactic fit, and the only observational data points plotted are those with the smallest uncertainties from the previous panel, i.e., those from {\it Spitzer} (filled blue squares) and {\it HST} (filled orange circle). Again, red lines connect the time of observation with its prediction. Pale purple solid and dashed lines mark lines of constant $\beta$ and $\lambda$, respectively, at 0$\farcs$1 spacing. (Lower left) RA and Dec as a function of time for the {\it Spitzer} and {\it HST} points. The best fit parallax curve as seen from the vantage points of {\it Spitzer} and {\it HST} is shown in blue and orange, respectively. The blue curve ends in January 2020, the end of the {\it Spitzer} mission. (Lower right) The RA and Dec residuals from the parallax fit as a function of time. Color coding is the same as in previous panels.
\label{fig:parallax_fit}}
\end{figure*}

This position was combined with our published unWISE and {\it Spitzer} astrometry and fed into the code of \cite{kirkpatrick2021} to produce a new astrometric solution (Figure~\ref{fig:parallax_fit}), resulting in an absolute parallax of 61.4$\pm$4.7 mas and a proper motion of 2687.1$\pm$11.3 mas yr$^{-1}$ (Table~\ref{tab:data}), the high value of the latter having been originally noted by \cite{meisner2020}.

%\startlongtable
\begin{deluxetable}{lll}
\tabletypesize{\scriptsize}
%\tablenum{1}
\tablecaption{Compiled Data for WISE 1534$-$1043\label{tab:data}}
\tablehead{
\colhead{Parameter} &
\colhead{Value\tablenotemark{a}} &                          
\colhead{Ref.}  \\
\colhead{(1)} &                          
\colhead{(2)} &
\colhead{(3)}
}
\startdata
AllWISE Designation& WISEA J153429.75$-$104303.3  & 2 \\
CatWISE Designation& CWISE J153429.19$-$104318.9  & 3 \\
\hline
\multicolumn{3}{c}{Astrometric Fit} \\
RA (ICRS) &  233.621873 deg $\pm$ 30.2 mas     & 1 \\
Dec (ICRS)&  $-$10.721775 deg $\pm$ 24.3 mas   & 1 \\
MJD       &  57645.36 d    & 1 \\
Epoch     &  2016.70 yr & 1 \\
$\varpi_{\rm abs}$& 61.4$\pm$4.7 mas & 1 \\
$\mu_{\rm RA}$ & $-$1253.1$\pm$8.9 mas yr$^{-1}$ & 1 \\
$\mu_{\rm Dec}$& $-$2377.0$\pm$7.0 mas yr$^{-1}$  & 1 \\
$\mu_{\rm total}$& 2687.1$\pm$11.3 mas yr$^{-1}$  & 1 \\
\hline
\multicolumn{3}{c}{Photometry} \\
$J_{\rm MKO}$  & $>$23.8 mag  & 1 \\
$H_{\rm MKO}$  & $>$18.58 mag  & 4,6 \\
$K_S$          & $>$17.85 mag  & 4,6 \\
F110W          & 24.695$\pm$0.083 mag  & 1 \\
W1             & 18.182$\pm$0.189 mag\tablenotemark{b}  & 3 \\
W2             & 16.145$\pm$0.084 mag\tablenotemark{b}  & 3 \\
W3             & $>$12.014 mag& 2 \\
W4             & $>$8.881 mag & 2 \\
ch1            & 16.691$\pm$0.032 mag & 4,5 \\
ch2            & 15.766$\pm$0.023 mag & 4,5 \\
$J-$W2         & $>$7.66 mag & 1 \\
$J-$ch2        & $>$8.03 mag & 1 \\
$F110W-$W2     & 8.550$\pm$0.118 mag & 1 \\
$F110W-$ch2    & 8.929$\pm$0.086 mag & 1 \\
W1$-$W2        & 2.037$\pm$0.207 mag & 3 \\
ch1$-$ch2      & 0.925$\pm$0.039 mag & 4,5\\
\hline
\multicolumn{3}{c}{Derived Quantities} \\
distance       & 16.3$^{+1.4}_{-1.2}$ pc & 1 \\
$v_{\rm tan}$  & 207.4$\pm$15.9 km s$^{-1}$  & 1 \\
$M_J$          & $>$22.7 mag  & 1 \\
$M_{F110W}$    & 23.636$\pm$0.19 mag  & 1 \\
$M_{W1}$       & 17.123$\pm$0.25 mag  & 1 \\
$M_{W2}$       & 15.086$\pm$0.19 mag  & 1 \\
$M_{ch1}$      & 15.632$\pm$0.17 mag  & 1 \\
$M_{ch2}$      & 14.707$\pm$0.17 mag  & 1 \\
\enddata
\tablecomments{Reference codes:
(1) This Letter,
(2) AllWISE Source Catalog - \citealt{cutri2013, kirkpatrick2014},
(3) CatWISE2020 Catalog - \citealt{marocco2021},
(4) \citealt{meisner2020},
(5) \citealt{kirkpatrick2021},
(6) VISTA Hemisphere Survey, DR6.
}
\tablenotetext{a}{Magnitudes are given in the Vega system. Conversions between Vega and AB magnitudes for {\it HST}/WFC3 filters can be found in {\url https://www.stsci.edu/hst/instrumentation/wfc3/data-analysis/photometric-calibration/ir-photometric-calibration.}}
\tablenotetext{b}{Magnitudes measured using the proper motion solution ($w1mpro\_pm$ and $w2mpro\_pm$), as these should be more accurate than those from the stationary solution.}
\end{deluxetable}

\section{Analysis\label{sec:analysis}}

In \cite{meisner2020}, WISE 1534$-$1043 was labeled as a color outlier on the $J-$ch2 vs.\ ch1$-$ch2 plot using a $J$-band magnitude limit of 20.56 mag from follow-up observations using Palomar/WIRC. The new $\it HST$ detection allows us to revisit this. 

\begin{figure*}
\includegraphics[scale=0.7,angle=0]{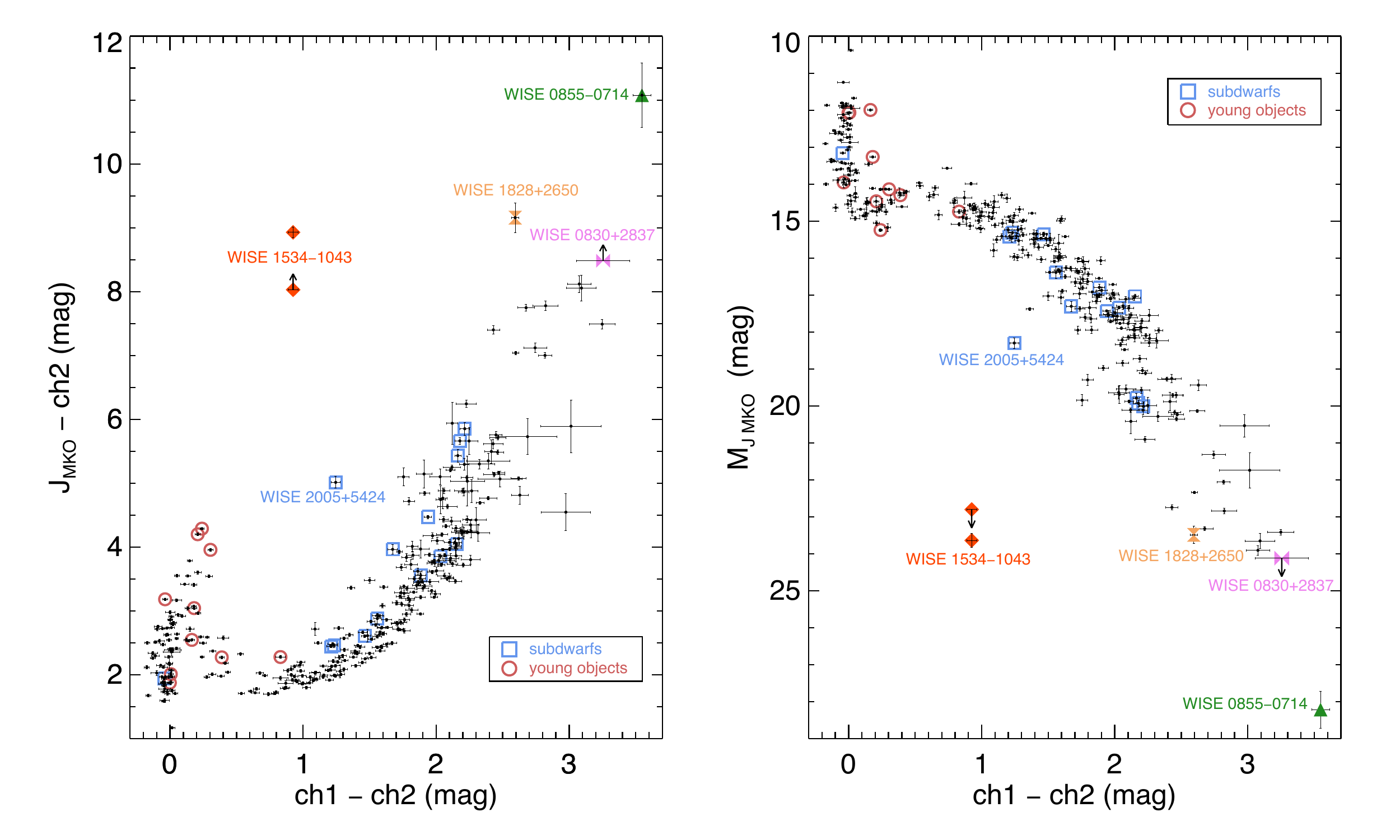}
\caption{Color-color and color-magnitude plots adapted from Figure 18 of \cite{kirkpatrick2021} and showing the 20-pc census of L, T, and Y dwarfs. (Left) $J_{\rm MKO}-$ch2 color vs.\ ch1$-$ch2 color. (Right) Absolute $J_{\rm MKO}$ magnitude vs.\ ch1$-$ch2 color. The locations of WISE 1534$-$1043 (orange diamond), WISE 0855$-$0714 (green triangle), WISE 0830+2837 (recumbent pink barbell), WISE 1828+2650 (standing yellow barbell), subdwarfs (blue squares), and young objects (ruddy circles) are marked along with other L, T, and Y dwarfs (black) within 20 pc of the Sun. There are two positions shown for WISE 1534$-$1043, as explained in the text. As a $J$-band limit for WISE 0830+2837, we have used four Y1 dwarfs with G141 spectra from \cite{schneider2015} to measure an offset of F125W$-J \approx 0.87$ mag, and we apply this offset to the F125W limit from \cite{bardalez2020}. The young objects and subdwarfs marked are those identified in Table 11 of \cite{kirkpatrick2021}. \label{fig:color_plots}}
\end{figure*}

If we assume that WISE 1534$-$1043 is a typical cold brown dwarf, we can  convert between F110W and $J$ magnitudes. Using synthetic photometry on thirteen high-S/N G102+G141 {\it HST} spectra of late-T and Y dwarfs from \cite{schneider2015}, we find that F110W magnitudes are fainter than $J$ magnitudes by 0.79 mag. This would imply that our F110W detection corresponds to $J \approx 23.9$ mag. However, this is only 0.1 mag fainter than the measured Keck $J$-band limit, meaning that the $Y$/$J$-band spectrum of WISE 1534$-$1043 likely differs from a typical late-T or Y dwarf. Because of this, we choose to plot two points for WISE 1534$-$1043 in Figure~\ref{fig:color_plots}. The first is the $J$-band limit from Keck, and the second is the detection from {\it HST}, for which we simply use the F110W magnitude as a proxy for $J$-band. As shown in  Figure~\ref{fig:color_plots} (left), WISE 1534$-$1043 is the sole occupant in its quadrant of the color-color diagram: it is bluer by $\sim$1.5 mag in ch1$-$ch2 color than any other object of similar $J-$ch2 color, and it is redder by at least 6 mag in $J-$ch2 than the majority of objects with similar ch1$-$ch2 color.

The unWISE- and {\it Spitzer}-based parallax in \cite{kirkpatrick2021} had a large (30\%) relative uncertainty. The addition of the {\it HST} data point now enables a parallax with a much smaller (7.7\%) relative uncertainty, placing the object securely within 20 pc. Comparing its absolute $J$ magnitude to those of other L, T, and Y dwarfs within 20 pc (Figure~\ref{fig:color_plots}, right), we find that the object is the sole occupant in its quadrant of the color-magnitude diagram. It is at least 8 mag fainter in $J$ than the bulk of objects of similar ch1$-$ch2 color, and the $M_J$ value is comparable to that of Y1 dwarfs (see Figure 16a of \citealt{kirkpatrick2021}).

\section{Possible Interpretations\label{sec:interpretation}}

The unique position of WISE 1534$-$1043 on these diagrams makes deduction of the object's nature challenging. Given that the object is too faint at all wavelengths for spectroscopy at any current facility, possible interpretations must rely on diagrams like Figure~\ref{fig:color_plots}, on which we try to recognize trends seen among other known objects and to extend these into a terra incognita guided by theoretical predictions. We consider four scenarios below.

\subsection {Extremely low-metallicity (old) brown dwarf} 

As discussed in \cite{meisner2020}, the color abnormalities in this object, even before a redder $J-$ch2 measurement was available, were reminiscent of those of WISE J200520.38+542433.9 (hereafter, WISE 2005+5424; \citealt{mace2013}). Shown as an outlying blue square on Figure~\ref{fig:color_plots}, WISE 2005+5424 is an sdT8 companion to Wolf 1130, which has a metallicity of [Fe/H] = $-0.64{\pm}0.17$ as measured from the M subdwarf primary (\citealt{rojas-ayala2012}) and a much bluer ch1$-$ch2 color (by $\sim$1 mag) than objects of comparable $J-$ch2 or $M_J$ values. 

Other {\it WISE} discoveries of unknown distance but occupying a similar locus of color space as WISE 2005+5424 also show spectroscropic indications of low metallicity (\citealt{schneider2020, meisner2021}). New atmospheric models, called LOWZ, are presented in \cite{meisner2021} and show that cold brown dwarfs tend to drift bluer in ch1$-$ch2 and redder in $J-$ch2 at a fixed temperature as metallicity decreases, which fits the trend expected if WISE 1534$-$1043 is a more metal-poor subdwarf than WISE 2005+5424. The bluer ch1$-$ch2 color may indicate a low relative abundance of methane, as expected in a low-Z atmosphere. Lower metallicity would also result in lower atmospheric opacity, meaning that the visible photosphere would extend deeper into the atmosphere where the pressure is higher. Because collision induced absorption by H$_2$ is predicted to be strongest in the near-infrared $JHK$ bands (\citealt{borysow1997}), this would suppress the $J$ flux and drive the $J-$ch2 color redward. 

The LOWZ models that best match the position of WISE 1534$-$1043 in the $J-$W2 vs.\ W1$-$W2 diagram suggest [M/H] $\approx -2.0$ and $T_{\rm eff} < 500$K, pushing it possibly into the Y dwarf sequence. If correct, this would make WISE 1534$-$1043 the first Y subdwarf. However, finding an object with a metallicity of $-2.0$ is extremely unusual at this distance. Based on a collection of the nearest, brightest K and M subdwarfs in the northern two-thirds of the sky (\citealt{kesseli2019}), we find that some objects in the 20-pc census (e.g., sdM3 LHS 272 and the aforementioned Wolf 1130) have metallicities near $-1$, but more metal-poor objects are unlikely to be found with 20 pc. At larger distances, for example, \cite{kesseli2019} measure a value of [Fe/H] = $-1.54{\pm}0.30$ for the usdM1 LHS 364, at 28 pc. It seems implausible, then, that WISE 1534$-$1043 would have a value as low as [M/H] = $-2$. Also, we note that the LOWZ models (Figure 7 of \citealt{meisner2021}) predict a value of $-1.5 < $[M/H] $< -1.0$ for WISE 2005+5424, which is much lower than the measured $-0.64{\pm}0.17$ value. Hence, the metallicity of WISE 1534$-$1043 is likely to be less extreme than the LOWZ models predict. 

\begin{figure}
\includegraphics[scale=0.4,angle=0]{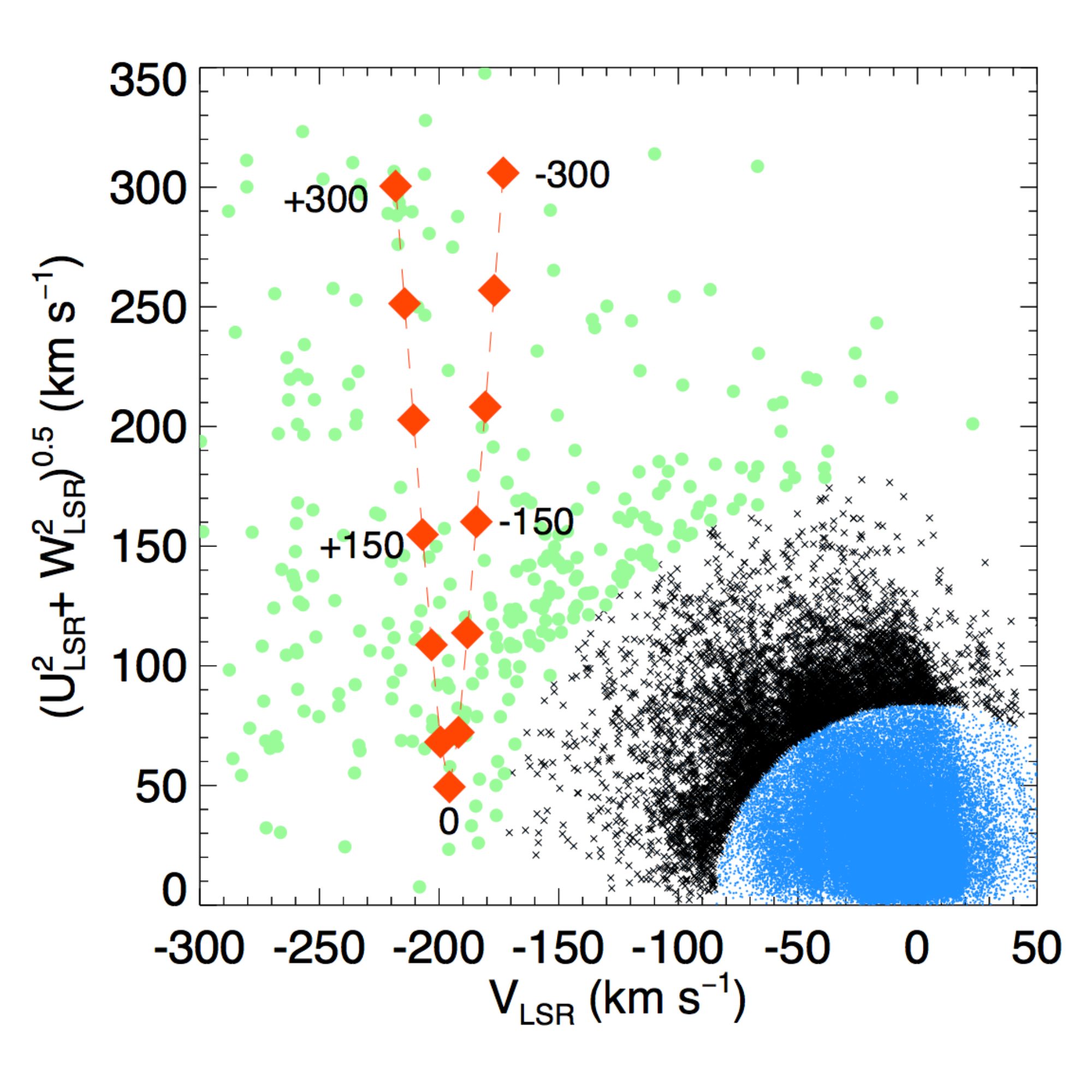}
\caption{Toomre diagram of $UVW$ space motions corrected to the Local Standard of Rest for 74,066 {\it Gaia} DR2 stars within 100 pc of the Sun and having parallax errors $< 10\%$. Thin disk (blue dots), thick disk (black crosses), and halo (green circles) objects are marked, per the text. WISE 1534$-$1043 (orange diamonds) is shown at thirteen radial velocities ranging from $-300$ to 300 km s$^{-1}$, in steps of 50 km s$^{-1}$. All values suggest a rare member of the halo.
\label{fig:UVW}}
\end{figure}

The low-metallicity hypothesis, which suggests old age, is also supported by the tangential velocity, 207.4$\pm$15.9 km s$^{-1}$ (Table~\ref{tab:data}), which exceeds by $\sim$50 km s$^{-1}$ the highest velocity known amongst the remainder of the 20-pc L, T, and Y dwarf census (Figure 33 of \citealt{kirkpatrick2021}). Figure~\ref{fig:UVW} is a Toomre diagram (\citealt{sandage1987}) showing the space motions of {\it Gaia} DR2 stars within the 100-pc volume and having well measured parallaxes. Using the kinematic criteria of \cite{nissen2004}, objects are coded as thin disk ($V_{\rm tot} \le 85$ km s$^{-1}$), thick disk ($85 < V_{\rm tot} \le 180$ km s$^{-1}$), or halo ($V_{\rm tot} > 180$ km s$^{-1}$). For all values of its (unknown) radial velocity, WISE 1534$-$1043 is suggested to be a halo member despite the fact that only 0.15\% of objects in the solar neighborhood are halo members (Table A1 of \citealt{bensby2014}).

\subsection {Extremely low-mass, low-gravity (young) brown dwarf} 

Young objects within the 20-pc sample are highlighted by red circles on Figure~\ref{fig:color_plots}. These show signs of low gravity in their spectra and/or are associated with young moving groups (see Section 7.1 of \citealt{kirkpatrick2021}), indicating that they have not yet contracted to their final, equilibrium radii (\citealt{faherty2016}). A small population with ch1$-$ch2 $<$ 0.5 mag have much redder $J-$ch2 colors than the others.  It is believed that in such objects, grains continue to amass in the photosphere as they cool because settling is unable to clear the upper atmosphere of condensates (\citealt{marley2012}) at the same temperatures where this occurs in the older, field dwarf population ($\sim$1250K; Figure 22 of \citealt{kirkpatrick2021}). As sufficiently cold $T_{\rm eff}$ values are reached, the rainout will eventually commence, pushing the L/T transition to a colder temperature. 

Under this scenario, WISE 1534$-$1043 would be a young brown dwarf that still has not crossed into the early-T dwarfs, although it would be $\sim$4.5 mag redder in $J-$ch2 color than any other young L dwarf yet known. The ch1$-$ch2 color of $\sim$1 mag indicates that a considerable amount of methane opacity is already present in the photosphere, which may be difficult to reconcile with a young, low-gravity hypothesis since the conversion of CO to CH$_4$ will also be pushed to lower values of T$_{\rm eff}$ (\citealt{zahnle2014}). Moreover, such a young object would not yet be kinematically heated and thus would be very unlikely to have a value of $v_{\tan}$ as large as that measured for WISE 1534$-$1043 (e.g., Figure 17 of \citealt{gonzales2019}).

\subsection {Ejected exoplanet} 

The two scenarios above assume WISE 1534$-$1043 is a brown dwarf formed via the star formation process. Instead, what if it is an exoplanet formed out of a protoplanetary disk and stripped later from its host system? \cite{beichman2013} posited this scenario as one possible explanation of the unusual features of WISE 1828+2650. The giant planets of our solar system are composed of material more metal rich than the Sun itself, showing that their formation process involves elemental segregation. Carbon, as one example, is enhanced in Jupiter by 3$\times$ the solar value (\citealt{owen1999}), in Saturn by 7$\pm$2$\times$ (\citealt{flasar2005}), and in Uranus and Neptune by 32$\times$ and 41$\times$, respectively (\citealt{lodders1994}). Under this hypothesis, WISE 1534$-$1043 is photometrically unusual because such elemental differences would profoundly affect its atmospheric composition and emergent spectrum. Unfortunately, forward modeling that incorporates a wide array of elemental abundance differences does not yet exist, so our best method to test this hypothesis is atmospheric retrieval, once a suitable spectrum for WISE 1534$-$1043 is obtained. 

The high $v_{\rm tan}$ measured for this object is not at odds with the ejected exoplanet hypothesis, but studies have found that the population of short timescale, single events observed in microlensing searches can be attributed to planets in wide orbits (\citealt{mroz2017, mroz2019}) rather than to a significant population of solivagant exoplanets. In addition, the "excess" short-timescale microlensing events -- those that cannot be attributed to the distribution of stars and brown dwarfs -- correspond to masses on the scale of 1 M$_{\rm Earth}$. Even if we assume that WISE 1534$-$1043 is an extremely young planet ($\sim$10 Myr) that has recently escaped and is still relatively bright, model predictions (\citealt{burrows1997}) suggest a mass of $\sim$0.3 M$_{\rm Jup}$ ($\sim$100 M$_{\rm Earth}$), far more massive than the bodies causing the short-timescale microlensing events. Assuming an older age only drives this mass upward. Having a rare, "high" mass, escaped planet falling within 20 pc of the Sun appears to be unlikely.

\subsection {Ultra-cold stellar remnant} 

Is it possible for WISE 1534$-$1043 to be a stellar remnant? Assuming a blackbody with a typical white dwarf radius of $\sim$7000 km would require $T_{\rm eff} \approx 1950$K to match the W2 flux and distance measured for WISE 1534$-$1043. However, a 1950K blackbody would have its peak flux near 1.5$\mu$m, which does not match the photometry of WISE 1534$-$1043. Although white dwarfs below $\sim$4000K have strong collision-induced absorption, reshaping their spectral energy distributions markedly away from a blackbody (\citealt{lenzuni1991}), this cannot explain the colors seen in WISE 1534$-$1043 (e.g., \citealt{blouin2017}.) Besides, theoretical models predict that it is not possible for a white dwarf to have cooled to this $T_{\rm eff}$ considering the age of the Milky Way (e.g., \citealt{calcaferro2018}). 

WISE 1534$-$1043 could be a more exotic, cold stellar remnant -- a bare core stripped through ablation (\citealt{ray2017}) and/or mass transfer (\citealt{hernandez2016}). For it to exist as an isolated object now, it would had to have been ripped from its ablating source through a gravitational encounter with a third object. Finding such a rare, ejected object within 20 pc of the Sun is highly improbable.

\section{Conclusions\label{sec:conclusion}}

We have better constrained the $J-$ch2 color, measured a robust parallax, and derived absolute magnitudes for WISE 1534$-$1043. These results were obtained through new $J$-band imaging from Keck that fails to detect the object and even deeper F110W imaging from {\it HST} that reveals the source at S/N $\approx$ 5. On the resultant color-color and color-magnitude diagrams, it falls in a region that contains no other known objects.

We conclude that the unique object WISE 1534$-$1043 is most likely a cold, very metal poor brown dwarf -- perhaps even the first Y-type subdwarf -- given its red $J-$ch2 color, its high tangential velocity, comparison to the photometric signatures seen in other low-metallicity objects, and trends suggested by theoretical predictions at low Z and low $T_{\rm eff}$. Verification, refutation, or further befuddlement should be possible via additional photometry and broad-wavelength spectroscopy from the {\it James Webb Space Telescope}.

\acknowledgments
This research is based on observations made under program 16243 with the NASA/ESA {\it Hubble Space Telescope} obtained from the Space Telescope Science Institute, which is operated by the Association of Universities for Research in Astronomy, Inc., under NASA contract NAS 5–26555. Some of the data presented here were obtained at the W.\ M.\ Keck Observatory, which is operated as a scientific partnership among the California Institute of Technology, the University of California, and the National Aeronautics and Space Administration. The Observatory was made possible by the generous financial support of the W.\ M.\ Keck Foundation. The authors wish to recognize and acknowledge the very significant cultural role and reverence that the summit of Maunakea has always had within the indigenous Hawaiian community.  We are most fortunate to have the opportunity to conduct observations from this mountain. This publication makes use of data products from {\it WISE/NEOWISE}, which is a joint project of UCLA and JPL/Caltech, funding by NASA. Portions of this research were carried out at JPL/Caltech under contract with NASA. JKF thanks the Heising Simons Foundation for research support.

\facilities{HST (WFC3), Keck:I (MOSFIRE), WISE, NEOWISE}

\software{WiseView (\citealt{caselden2018}).}


\begin{thebibliography}{}
\bibitem[Bardalez Gagliuffi et al.(2020)]{bardalez2020} Bardalez Gagliuffi, D.~C., Faherty, J.~K., Schneider, A.~C., et al.\ 2020, \apj, 895, 145. doi:10.3847/1538-4357/ab8d25
\bibitem[Bedin \& Fontanive(2018)]{bedin2018} Bedin, L.~R. \& Fontanive, C.\ 2018, \mnras, 481, 5339. doi:10.1093/mnras/sty2626
\bibitem[Beichman et al.(2013)]{beichman2013} Beichman, C., Gelino, C.~R., Kirkpatrick, J.~D., et al.\ 2013, \apj, 764, 101. doi:10.1088/0004-637X/764/1/101
\bibitem[Bensby et al.(2014)]{bensby2014} Bensby, T., Feltzing, S., \& Oey, M.~S.\ 2014, \aap, 562, A71. doi:10.1051/0004-6361/201322631
\bibitem[Blouin et al.(2017)]{blouin2017} Blouin, S., Kowalski, P.~M., \& Dufour, P.\ 2017, \apj, 848, 36. doi:10.3847/1538-4357/aa8ad6
\bibitem[Borysow et al.(1997)]{borysow1997} Borysow, A., Jorgensen, U.~G., \& Zheng, C.\ 1997, \aap, 324, 185
\bibitem[Burrows et al.(1997)]{burrows1997} Burrows, A., Marley, M., Hubbard, W.~B., et al.\ 1997, \apj, 491, 856. doi:10.1086/305002
\bibitem[Calcaferro et al.(2018)]{calcaferro2018} Calcaferro, L.~M., Althaus, L.~G., \& C{\'o}rsico, A.~H.\ 2018, \aap, 614, A49. doi:10.1051/0004-6361/201732551
\bibitem[Caselden et al.(2018)]{caselden2018} Caselden, D., Westin, P., Meisner, A., et al.\ 2018, Astrophysics Source Code Library. ascl:1806.004
\bibitem[Cushing et al.(2021)]{cushing2021} Cushing, M.~C., et al.\ 2021, submitted
\bibitem[Cushing et al.(2011)]{cushing2011} Cushing, M.~C., Kirkpatrick, J.~D., Gelino, C.~R., et al.\ 2011, \apj, 743, 50. doi:10.1088/0004-637X/743/1/50
\bibitem[Cutri et al.(2013)]{cutri2013} Cutri, R.~M., Wright, E.~L., Conrow, T., et al.\ 2013, Explanatory Supplement to the AllWISE Data Release Products, by R. M. Cutri et al.
\bibitem[Faherty et al.(2016)]{faherty2016} Faherty, J.~K., Riedel, A.~R., Cruz, K.~L., et al.\ 2016, \apjs, 225, 10. doi:10.3847/0067-0049/225/1/10
\bibitem[Flasar et al.(2005)]{flasar2005} Flasar, F.~M., Achterberg, R.~K., Conrath, B.~J., et al.\ 2005, Science, 307, 1247. doi:10.1126/science.1105806
\bibitem[Gonzales et al.(2019)]{gonzales2019} Gonzales, E.~C., Faherty, J.~K., Gagn{\'e}, J., et al.\ 2019, \apj, 886, 131. doi:10.3847/1538-4357/ab48fc
\bibitem[Hern{\'a}ndez Santisteban et al.(2016)]{hernandez2016} Hern{\'a}ndez Santisteban, J.~V., Knigge, C., Littlefair, S.~P., et al.\ 2016, \nat, 533, 366. doi:10.1038/nature17952
\bibitem[Kesseli et al.(2019)]{kesseli2019} Kesseli, A.~Y., Kirkpatrick, J.~D., Fajardo-Acosta, S.~B., et al.\ 2019, \aj, 157, 63. doi:10.3847/1538-3881/aae982
\bibitem[Kirkpatrick et al.(2021)]{kirkpatrick2021} Kirkpatrick, J.~D., Gelino, C.~R., Faherty, J.~K., et al.\ 2021, \apjs, 253, 7. doi:10.3847/1538-4365/abd107
\bibitem[Kirkpatrick et al.(2014)]{kirkpatrick2014} Kirkpatrick, J.~D., Schneider, A., Fajardo-Acosta, S., et al.\ 2014, \apj, 783, 122. doi:10.1088/0004-637X/783/2/122
\bibitem[Lang(2014)]{lang2014} Lang, D.\ 2014, \aj, 147, 108. doi:10.1088/0004-6256/147/5/108
\bibitem[Lenzuni et al.(1991)]{lenzuni1991} Lenzuni, P., Chernoff, D.~F., \& Salpeter, E.~E.\ 1991, \apjs, 76, 759. doi:10.1086/191580
\bibitem[Lodders \& Fegley(1994)]{lodders1994}
Lodders, K. \& Fegley, B.\ 1994, \icarus, 112, 368. doi:10.1006/icar.1994.1190
\bibitem[Luhman(2014)]{luhman2014} Luhman, K.~L.\ 2014, \apjl, 786, L18. doi:10.1088/2041-8205/786/2/L18
\bibitem[Mace et al.(2013)]{mace2013} Mace, G.~N., Kirkpatrick, J.~D., Cushing, M.~C., et al.\ 2013, \apj, 777, 36. doi:10.1088/0004-637X/777/1/36
\bibitem[Mainzer et al.(2014)]{mainzer2014} Mainzer, A., Bauer, J., Cutri, R.~M., et al.\ 2014, \apj, 792, 30. doi:10.1088/0004-637X/792/1/30
\bibitem[Marley et al.(2012)]{marley2012} Marley, M.~S., Saumon, D., Cushing, M., et al.\ 2012, \apj, 754, 135. doi:10.1088/0004-637X/754/2/135
\bibitem[Marocco et al.(2021)]{marocco2021} Marocco, F., Eisenhardt, P.~R.~M., Fowler, J.~W., et al.\ 2021, \apjs, 253, 8. doi:10.3847/1538-4365/abd805
\bibitem[McLean et al.(2012)]{mclean2012} McLean, I.~S., Steidel, C.~C., Epps, H.~W., et al.\ 2012, \procspie, 8446, 84460J. doi:10.1117/12.924794
\bibitem[McLean et al.(2010)]{mclean2010} McLean, I.~S., Steidel, C.~C., Epps, H., et al.\ 2010, \procspie, 7735, 77351E. doi:10.1117/12.856715
\bibitem[Meisner et al.(2021)]{meisner2021} Meisner, A.~M., et al.\ 2021, in prep.
\bibitem[Meisner et al.(2020)]{meisner2020} Meisner, A.~M., Caselden, D., Kirkpatrick, J.~D., et al.\ 2020, \apj, 889, 74. doi:10.3847/1538-4357/ab6215
\bibitem[Meisner et al.(2018a)]{meisner2018a} Meisner, A.~M., Lang, D., \& Schlegel, D.~J.\ 2018, \aj, 156, 69. doi:10.3847/1538-3881/aacbcd
\bibitem[Meisner et al.(2018b)]{meisner2018b} Meisner, A.~M., Lang, D.~A., \& Schlegel, D.~J.\ 2018, Research Notes of the American Astronomical Society, 2, 202. doi:10.3847/2515-5172/aaecd5
\bibitem[Mr{\'o}z et al.(2019)]{mroz2019} Mr{\'o}z, P., Udalski, A., Skowron, J., et al.\ 2019, \apjs, 244, 29. doi:10.3847/1538-4365/ab426b
\bibitem[Mr{\'o}z et al.(2017)]{mroz2017} Mr{\'o}z, P., Udalski, A., Skowron, J., et al.\ 2017, \nat, 548, 183. doi:10.1038/nature23276
\bibitem[Nissen(2004)]{nissen2004} Nissen, P.~E.\ 2004, Origin and Evolution of the Elements, 154
\bibitem[Owen et al.(1999)]{owen1999} Owen, T., Mahaffy, P., Niemann, H.~B., et al.\ 1999, \nat, 402, 269. doi:10.1038/46232
\bibitem[Ray \& Loeb(2017)]{ray2017} Ray, A. \& Loeb, A.\ 2017, \apj, 836, 135. doi:10.3847/1538-4357/aa5b7d
\bibitem[Renedo et al.(2010)]{renedo2010} Renedo, I., Althaus, L.~G., Miller Bertolami, M.~M., et al.\ 2010, \apj, 717, 183. doi:10.1088/0004-637X/717/1/183
\bibitem[Rojas-Ayala et al.(2012)]{rojas-ayala2012} Rojas-Ayala, B., Covey, K.~R., Muirhead, P.~S., et al.\ 2012, \apj, 748, 93. doi:10.1088/0004-637X/748/2/93
\bibitem[Sandage \& Fouts(1987)]{sandage1987} Sandage, A. \& Fouts, G.\ 1987, \aj, 93, 74. doi:10.1086/114291
\bibitem[Schneider et al.(2020)]{schneider2020} Schneider, A.~C., Burgasser, A.~J., Gerasimov, R., et al.\ 2020, \apj, 898, 77. doi:10.3847/1538-4357/ab9a40
\bibitem[Schneider et al.(2015)]{schneider2015} Schneider, A.~C., Cushing, M.~C., Kirkpatrick, J.~D., et al.\ 2015, \apj, 804, 92. doi:10.1088/0004-637X/804/2/92
\bibitem[Wright et al.(2014)]{wright2014} Wright, E.~L., Mainzer, A., Kirkpatrick, J.~D., et al.\ 2014, \aj, 148, 82. doi:10.1088/0004-6256/148/5/82
\bibitem[Wright et al.(2010)]{wright2010} Wright, E.~L., Eisenhardt, P.~R.~M., Mainzer, A.~K., et al.\ 2010, \aj, 140, 1868. doi:10.1088/0004-6256/140/6/1868
\bibitem[Zahnle \& Marley(2014)]{zahnle2014} Zahnle, K.~J. \& Marley, M.~S.\ 2014, \apj, 797, 41. doi:10.1088/0004-637X/797/1/41
\end{thebibliography}
\end{document}